\documentclass{elsart}

\usepackage{graphicx}
\usepackage{amssymb}
\newcommand{\gtsim}{\mathop{\,>\kern-1.05em\lower1.ex\hbox{$\sim$}\,}}

\begin{document}

\begin{frontmatter}



\title{Model study for the nonequilibrium magnetic domain structure 
during the growth of nanostructured ultrathin films}


\author{R. Brinzanik\corauthref{cor1}},
\corauth[cor1]{Corresponding author. Tel.: +49-30-83854784; fax.: +49-30-83856799}
\ead{brinzani@physik.fu-berlin.de}
\author{P.J. Jensen},
\author{K.H. Bennemann}

\address{Institut f\"ur Theoretische Physik, Freie Universit\"at 
Berlin,\\ Arnimallee 14, D-14195 Berlin, Germany}

\begin{abstract}
The  nonequilibrium  magnetic domain  structure  of growing  ultrathin
ferromagnetic films with a realistic  atomic structure is studied as a
function of coverage and temperature.   We apply a kinetic Monte Carlo
method  to  a  micromagnetic  model  describing  the  transition  from
superparamagnetic islands  at low coverages to  a closed ferromagnetic
film.   The   magnetic  relaxation   and  the  island   growth  happen
simultaneously.  Near the  percolation threshold a metastable magnetic
domain  structure is  obtained  with an  average  domain area  ranging
between the  area of individual magnetic  islands and the  area of the
large domains  observed for thicker ferromagnetic  films.  We conclude
that this  micro-domain structure is controlled and  stabilized by the
nonuniform  atomic  nanostructure of  the  ultrathin  film, causing  a
random  interaction between  magnetic islands  with varying  sizes and
shapes. The average domain area and domain roughness are determined. A
maximum of the  domain area and a minimum of  the domain roughness are
obtained as a function of the temperature.
\end{abstract}

\end{frontmatter}

% main text
\section{Introduction}
\label{Introduction} 
The investigation of the magnetic domain formation, in particular in 
nano\-structured ultrathin films, is an active field of current research. 
The influence of the atomic morphology on the magnetic properties is known to be 
especially strong during the initial states of the thin film growth. 
A small variation of the preparation
conditions may change the corresponding magnetic structure markedly. 
This has been shown by the recent progress of 
highly resolving imaging techniques, which allows for the investigation of 
the atomic structure as well as of the magnetic domain structure of these 
nanostructured systems in greater 
detail \cite{ASB90,JSO95,OSMK97,MBW97,KGK00,KPB00}. For example, a strong 
dependence of the domain structure on the film morphology has been observed 
for the Co/Au(111) thin film system yielding a smaller average domain 
area for rougher Co films \cite{ASB90,JSO95}. In addition, we 
emphasize that the consideration of nonequilibrium states for the investigation 
of the magnetic domain structure is of particular importance. 

In this contribution we study the interplay of the atomic and magnetic 
structure theoretically. In particular we simulate 
the  spatial and  temporal  magnetic domain  formation   
during the {\em simultaneous} growth of ultrathin ferromagnetic films. 
To our knowledge this problem
has not been investigated previously. The domain 
structure is calculated  within a kinetic Monte  Carlo (KMC) simulation 
applied to a micromagnetic model considering 
exchange coupling and uniaxial lattice anisotropy.  
The  average domain  area $\overline  S$ and  the average  domain roughness
$\overline R$ of  the growing thin film are  calculated as functions of
coverage $\Theta$ and temperature $T$. The domain roughness is
defined as the edge-to-area ratio  of a domain. 

In general, for low coverages
the  growing  thin  film mostly consists  of  isolated 
superparamagnetic islands with a domain area of
the order of  the island area. On the other hand, at higher coverages
the coagulation  of islands results in  a
closed ferromagnetic film which 
allows for the formation of large domains  with sizes 
of the  order  of about $1\,\mu$m, as usually observed \cite{ASB90}. 
The crossover between these two extremal domain size ranges
is still not well investigated. Also, the quantities determining
this transition are not known.

For the simulation of the magnetic properties of the growing thin film we 
consider the following important features:

i) We take into account an irregular, nanostructured film morphology 
which is usually present in a realistic thin film system. This is performed 
with the help of a simple growth model, allowing for the 
consideration of a size range between the 
atomic and the $\mu$m-scale. The main quantities determining the growth 
(island density and -arrangement, growth mode, etc.) 
are chosen in accordance with experiments. Due to the strong direct exchange 
coupling each island is assumed to be single-domain, 
its magnetization rotates thus coherently \cite{StW48}. 

ii) In  the  coverage  range  near  the  percolation threshold $\Theta_\mathrm{P}$ 
of  the thin  film the  interplay between  the (irregular) atomic
morphology  and  the  magnetic  domain  structure is  expected  to  be
strongest. Near $\Theta_\mathrm{P}$ the domain formation is governed by
'weak links' between the magnetic islands, while the island-island interaction 
tries to reach a ferromagnetic (single-domain) state of the film. 
The islands of a homogeneous, periodic island ensemble merge at 
the same time, thus the crossover from small to large domains takes place 
simultaneously all over the growing film. The characteristic times to 
cross the energy barriers, depending on the anisotropy and the island size, 
are the same for all islands. For an irregular nanostructure, however, 
the interactions between different island pairs will be strongly 
nonuniform (random magnet), causing various energy barriers for the 
magnetic rotation of different islands, and various 
inherent characteristic times \cite{ChB94,OhC86}. 
A pre-formation of domains with 
intermediate areas covering several neighbouring islands is expected. 
This property infers much longer magnetic relaxation  times to reach the thermal 
equilibrium than for a periodic island array. In other 
words, such a random ferromagnetic system can be more easily trapped into 
metastable states with a frozen-in multi-domain structure, and with an average domain 
size in between the above mentioned extremal sizes. 
Here the consideration of a realistic irregular atomic structure is 
especially important. 

iii) A finite temperature is taken into account in order to investigate 
the domain formation of a realistic thin film system. The temperature 
affects strongly the probabilities of the island magnetizations to overcome 
the energy barriers resulting in a long range magnetic order depending 
on time and temperature. In addition, also the {\em internal\/} (short range) 
magnetic order of a single island due to the finite exchange coupling is considered, 
leading to temperature dependend island-island interactions and lattice 
anisotropies. 

iv) Furthermore, we emphasize that for the  investigation of
such an ensemble of magnetic islands the consideration  
of states far from thermal equilibrium is very  important. 
The long relaxation  times common to a random magnet may 
be much larger than typical detecting times.  
We expect in particular that in the 
coverage range around $\Theta_\mathrm{P}$ the domain structure with an
intermediate average domain area may be observable within the 
measuring time or the simulation time. 

v) Most importantly, the   equilibrium   state   also   changes
permanently, if during the magnetic relaxation the thin film
grows steadily. Thus, within our simulation of the magnetic domain 
stucture we take simultaneously into account the temporal 
variation of the atomic structure during the thin film growth. This results
consequently in a temporal variation of the magnetic interactions 
determined by the growth velocity. 

The formation of large magnetic  domains is expected to be facilitated
by an increasing  temperature, since then energy barriers  can be more
easily surmounted.  This  will result in a larger average domain area
$\overline  S$ after a given observation 
time, or  a faster  formation of larger  domains.  On the  other hand,
fluctuations  at elevated  temperatures  will  destroy  large  domains
yielding a  smaller average domain area.  Thus, a maximum of 
$\overline{S}(T)$ as a function of temperature during domain formation 
is expected. With the same 
argument we expect a minimum for the average domain roughness 
$\overline{R}(T)$. 

Previous simulations using similar micromagnetic models 
have   been   performed  mainly   for
the investigation of the domain structure 
during the magnetization reversal of   thicker
ferromagnetic films  with a granular structure  \cite{Man87}.  
In these studies the systems have been divided  into periodic
arrays of identical magnetic cells
with a  nonuniform distribution  of the magnetic  interactions between
the  cells. In contrast, within our model we consider an {\em irregular 
atomic structure}. Different anisotropies and magnetic island-island 
interactions are caused by the island size dispersion and different 
common surface areas between neighboring islands.  
To  our knowledge such a realistic inhomogeneous atomic structure
of an ultrathin film in the early states of film growth for the 
investigation of the spatial domain formation
has not been considered previously.
  
In addition, theoretical  investigations often consider equilibrium  
magnetic  structures \cite{footnote1}. As mentioned, the relaxation times 
to reach the equilibrium states may be very long, so that in experiments 
metastable multi-domain structures are usually observed.  In the 
present study we will simulate such domain patterns by calculating the 
temporal variation of the magnetic structure far from equilibrium.
Since analytical approaches are not available for the study 
of the domain formation in a large system far from thermal equilibrium, 
we are forced to apply a kinetic (time resolved) Monte Carlo
simulation. 

This work is structured as follows.  In Sec.~\ref{Model}  we outline our model. 
A simple  growth model is applied which allows for the consideration
of large  systems. Magnetic interactions  are taken into account  by a
micromagnetic model, the magnetic relaxation is determined within a kinetic
Monte Carlo method.  Results for the average domain area and domain 
roughness as functions of coverage and temperature are presented in 
Sec.~\ref{Results}. A conclusion is given in Sec.~\ref{Conclusion}.

\section{MODEL}\label{Model} 
{\bf Eden Growth Model:} 
We apply  a Monte Carlo (MC)  method \cite{Bin79} to simulate the  molecular beam
epitaxial (MBE) growth  of a magnetic ultrathin film  on a nonmagnetic
substrate  using a  simple  solid-on-solid growth  model. Within  this
so-called  Eden model  \cite{Bin79,Ede58}  each atom  is  randomly attached  to
already existing islands, and stays immobile afterwards. To take into 
account different growth modes we assume furthermore that the  adatoms
are placed on lattice sites $i$  with varying probabilities
\begin{equation}   
p(q_i,z)   \propto \exp\left(- A(z)\,\sqrt{q_i}\,\right) \;. \label{e1}
\end{equation}
Here $z$ refers to the layer index. The  square-root dependence on  the  local 
coordination numbers $q_i$ has been
obtained to  be approximately  valid for metal  surfaces \cite{MHS92}.
By using  layer dependent  binding parameters $A(z)$  we are  able to
simulate with  simple  means different growth modes such as an 
island-type growth    mode   (three-dimensional   islands) or    a   
layer-by-layer growth  mode. 
Also various  surface faces and island  densities and -arrangements can  be 
considered  in accordance  with experiments,  for example islands 
arranged in chains. 
This  modified Eden  growth model  is  valid for  the case  of a  fast
surface  diffusion  and  a  moderate  step diffusion  for  which  the
mobility of  the atoms is large  enough to probe  the different atomic
positions of the island edges.  Since the very time consuming
calculation  of the  atomic diffusion is avoided,  we are  able to  
consider a  large number  of  nonequivalent sites ($\sim\,10^6$). 
The  influence of the temperature on the  thin film
morphology is not considered within our study \cite{footnote2}. 

In  the present  calculations we  assume a  $500\times500$  atomic and
magnetic unit  cell ($2.5\cdot10^5$ sites) on a  fcc(001) lattice with
periodic  boundary  conditions. We choose as an example the
bilayer  growth mode of the  first two  atomic layers  as observed  for the
Co/Cu(001) thin film system \cite{ScK92}, using always the ratio of 
the binding parameters $A(1)/A(2)=0.989$. 
In accordance with  experiment the island
density is  put equal  to 0.005 per  lattice site, refering  to $1250$
elementary islands in the unit cell. The simulation is started with
a random distribution of occupied  sites with minimal mutual distance 
$r_{\mathrm{min}}=10$ lattice constants which serve as seeds. 
Each additional atom
is placed on a perimeter site with probability $p(q_i,z)$. 
We stop the  growth after deposition
of two magnetic layers. The full coverage represents a system with two
completely filled  layers. 

{\bf Magnetic Structure:} 
For a given atomic structure during the growth of the thin film the 
corresponding magnetic properties are determined by performing 
a kinetic Monte Carlo simulation \cite{Bin79}. The 
following micromagnetic model for the total energy of
a system of interacting magnetic islands is assumed: 
\begin{equation}
E=-\frac{1}{4} \sum_{i,j}\gamma_{ij}(T,\Theta)\;L_{ij}(\Theta)\; \vec{S}_i\;
\vec{S}_j\; - \sum_i K_i(T,\Theta)\;N_i(\Theta)\;(S_i^z)^2 \;, \label{e2}
\end{equation}  
where $\vec{S}_i$ is a normalized vector characterizing the 
direction of the magnetization of the $i$-th island, 
$T$ is the temperature, and $\Theta$ the coverage of the film. 
Each magnetic island with $N_i$ atoms is assumed to be single-domain 
(Stoner-Wohlfahrt particle \cite{StW48}) with a single
giant magnetic moment $\mu_i=N_i\; \mu_{\mathrm{at}}$, $\mu_{\mathrm{at}}$ 
the atomic magnetic moment.  
The island moments are subject to a uniaxial lattice anisotropy $K_i$ 
per atomic spin. Due to this anisotropy we simplify our calculations 
by considering only two different directions along the easy axis 
for each island moment ($S_i=\pm\,1$).
This corresponds to the 
moments directed 'up' and 'down' for a perpendicular film magnetization,
or to the two orientations of the in-plane magnetization in case of 
a rectangular (110) layer, for example. 
For an isolated island these two states are separated by an 
energy barrier given by its total lattice anisotropy
$\Delta E_i=K_i\,N_i$ which has to be surmounted during 
magnetic reversal \cite{footnote3}. 
During the thin film growth two islands $i$ and $j$ may merge. 
If the directions of their magnetizations are different, 
the two islands minimize their mutual exchange coupling by the creation of 
a magnetic domain wall with the 
energy $E=\gamma_{ij}\,L_{ij}$. Here $\gamma_{ij}$ is the domain wall energy
per bond and $L_{ij}(\Theta)$ is the
common surface area of the islands in units of the 
lattice constant a$_o$ \cite{footnote4}. 
For simplicity the magnetic dipole coupling is not considered here.

We emphasize that the energy expression Eq. (\ref{e2}), 
representing a system of individual magnetic islands with varying interactions 
$\gamma_{ij}\,L_{ij}$, is a good approximation for nanostructured systems. 
By application of this model we can describe isolated superparamagnetic 
islands at low coverages ($L_{ij}(\Theta)=0$), as well as 
a strongly connected ferromagnetic film at high coverages, 
and the transition between these two extremes during the thin film growth. 
However, in the case of a smooth film (in the present study for 
$\Theta\gtsim 1.8$) the maintenance of irregularly shaped 
magnetic islands with nonuniform couplings is an unphysical discretization 
of the system. Thus in such a coverage range our model is not valid. 

Furthermore, the decreasing {\em internal magnetic order\/} 
of the islands with increasing temperature results 
in temperature dependend island interactions $\gamma_{ij}(T)$ and
{\em effective anisotropy coefficients} $K_i(T)$. We treat
the atomic magnetic moments as $S=1$ quantum spins. Within a
mean field theory the relative internal island magnetization $m_i(T,\Theta)$ 
is given by the Brillouin function 
\begin{eqnarray}
m_i(T,\Theta)=S\,B_1(x_i)&=&\frac{3}{2}\coth\left(\frac{3}{2}\,x_i\right)
-\frac{1}{2}\coth\left(\frac{1}{2}\,x_i\right) \;, \label{e3} \\[0.3cm] 
x_i&=&\frac{\overline{z}_i(\Theta)\;J\;m_i(T,\Theta)}{k_\mathrm{B}\,T} \;. \label{e4}
\end{eqnarray} 
Here, $\overline{z}_i(\Theta)$ is the average coordination 
number of island $i$ which depends on the coverage $\Theta$ of the
growing thin film, $k_\mathrm{B}$ is the Boltzmann constant. 
The ability of the anisotropy to maintain a certain
direction of the magnetization decreases due to thermal agitation.
Thus, a decreasing 
$m_i(T,\Theta)$ causes also a decreasing $K_i(T)$. Within a first order 
thermodynamic pertubation theory \cite{MF95} the 
effective anisotropy  $K_i(T)$ for $S=1$ is given by 
\begin{eqnarray}
K_i(T,\Theta) & = & K\;f_i(T,\Theta )\nonumber\\
              & = & K\;\bigg[4-
\frac{9}{2}\,\coth\left(\frac{3}{2}\,x_i\right)
\,\coth\left(\frac{1}{2}\,x_i\right)+
\frac{3}{2}\,\coth^2\left(\frac{1}{2}\,x_i\right)\bigg]\;. \label{e5}
\end{eqnarray}   
Note that $m_i(T)\,\to\,1$ and $f_i(T)\,\to\,1$ for $T\,\to\,0$.
Following the general derivation of the domain wall width \cite{Hub98}, 
the influence of the finite temperature on the interaction between the 
islands is considered as 
\begin{equation}
\gamma_{ij}(T,\Theta) = \gamma\;\sqrt{m_i(T,\Theta)\;m_j(T,\Theta)\;
f_{ij}(T,\Theta)} \;, \label{e6}  
\end{equation} 
where $f_{ij}(T,\Theta)$ is the average of $f_i(T,\Theta)$ and $f_j(T,\Theta)$, 
see Eq. (\ref{e5}), of islands $i$ and $j$. 

The temporal development of the magnetic arrangement of the island 
ensemble is determined as follows. Thermal activation and
(surface) interaction energy between neighboring islands may cause
reversals of the island magnetizations between the two states $S_i=\pm\,1$.
This relaxational behavior during the 
film growth is calculated with the help of the KMC method \cite{Bin79}.
From Eq. (\ref{e2}) the energy of island $i$ as function of the angle $\phi$ 
between $\vec{S}_i$ and the easy axis is given by
\begin{equation}
\epsilon_i(\phi)=E_i/K_i\,N_i=-2 h_i \cos \phi - \cos^2 \phi \;, \label{e7}
\end{equation}
with $h_i=S_i \sum_j \gamma_{ij}\,L_{ij}\,S_j/4K_i\,N_i$. Here we have made use 
of the condition $S_j = \pm 1$. By analysing Eq. (\ref{e7}) two cases have to
be distinguished. 

First for $|h_i| \leq 1$ the two states $S_i = \pm 1$ ($\phi = 0, \pi$) both 
represent energy minima. These are separated by an energy barrier 
governed by the competing volume energy (lattice anisotropy) and 
surface energy contributions. The barriers for the transitions 
$(S_i=+1) \rightleftharpoons (S_i=-1)$ are given by 
$\Delta E_i=N_i\,K_i (h_i\pm1)^2$.  
The flip rate $\Gamma_i$ of island $i$ 
to overcome $\Delta E_i$ is calculated by use of the 
Arrhenius-type ansatz \cite{Nee49}
\begin{equation}
\Gamma_i=\Gamma_o \exp\left(\frac{-\Delta E_i}
   {k_\mathrm{B} T}\right) \;, \label{e8}
\end{equation}
where the prefactor $\Gamma_o$ is treated as a constant 
'attempt frequency', determining the time unit of the 
magnetic relaxation.

Secondly, for $|h_i|>1$ one of the states 
$S_i=\pm1$ refers to an energy maximum and the other to a minimum. 
Then  $\Delta E_i=\pm 4 N_i\,K_i\, h_i$ is the energy difference 
between the two directions of the island magnetization and is governed only by
the surface energy.
This case is treated with the usual Metropolis algorithm
\cite{Bin79,Metro}, using the same prefactor $\Gamma_o$ as in Eq. (\ref{e8}). 

The growing thin film is characterized by 
a large amount of nonequivalent lattice sites, corresponding to 
a large number of different interaction parameters. 
Since little is known about these values, we use in our
simulation averaged quantities for $\gamma$, $J$, and $K$ which are fixed as
follows, using as an example the Co/Cu(001) thin film system. 
The domain wall energy $\gamma$ is adjusted to 
give the observed Curie temperature of the ferromagnetic {\em long range\/} order
of $T_\mathrm{C}=350$~K of a Co/Cu(001) film with two monolayers (ML) 
\cite{BPP99}. We obtain $\gamma=5.8$~meV per bond \cite{footnote4}.  
In a nanostructured film consisting of connected islands the {\em internal\/} 
ordering temperature of the islands $T_\mathrm{C}^{\mathrm{int}}$ 
is larger than $T_\mathrm{C}$.
Due to the lack of measurements for $T_\mathrm{C}^{\mathrm{int}}$, 
the exchange coupling $J$ 
is chosen to yield $T_\mathrm{C}^{\mathrm{int}}=450$~K at a coverage of 
2~ML, refering to a coordination number $z=8$. This results in $J=7.3$~meV
per nearest neighbor bond. The uniaxial anisotropy constant is set equal to 
$K=0.2$~meV per atom, which is a typical value for ultrathin films \cite{BlH94}. 
For comparison, with these values the width of an  {\em undisturbed\/} 
Bloch wall in a bulk ferromagnet is given by 
$w_\mathrm{B}=\pi\,\sqrt{J/2K}\approx$ 13~a$_o$, and the 
corresponding wall energy $\gamma_\mathrm{B}=4\sqrt{JK/2}=3.4$~meV per bond \cite{Hub98}.
The 'attempt frequency' $\Gamma_o$ is set equal to 
$\Gamma_o=5.0\cdot10^9$~sec$^{-1}$ \cite{Nee49}. The growth velocity of the film
is adjusted to 1~ML/100~sec.

The simulation procedure is performed as follows: first the island seeds
with random spin directions are distributed over the unit cell with periodic
boundary conditions. 
Then for a given temperature the thin film grows  
up to a coverage of 2~ML, which is divided 
into 400 growth steps. After each growth step the quantities 
$N_i(\Theta)$, $L_{ij}(\Theta)$, $\gamma_{ij}(T,\Theta)$, and 
$K_i(T,\Theta)$ are calculated. By applying 
100 Monte Carlo steps (MCS) per island moment of randomly chosen islands 
the magnetic structure of the island ensemble 
is allowed to relax towards its thermal equilibrium. Then the growth 
procedure is repeated.
 
The magnetic domains are determined by identifying connected islands 
with a parallel magnetization (Hoshen-Kopelman algorithm \cite{HoK76}). 
From this procedure the number of domains as well as their 
average area $\overline S$ and average roughness
$\overline  R$ are calculated.
Only  the first magnetic layer is used  for this analysis.
To indicate the influence of the atomic structure on the magnetic domain
properties, we calculate also the average area of the elementary islands,
as well as the average area of connected islands, consisting of coagulated 
elementary islands.    

The  domain roughness $R_i$  is defined as
the ratio $R_i =\partial S_i/S_i$ of  the number of 
atomic magnetic moments $\partial S_i$ 
at the  outer edge of the domain and its total number $S_i$. Since
for a domain with an increasing area its edge-to-area ratio is always
decreasing, we consider here rather the average {\em relative domain roughness\/}
$\overline r = \overline{R/R_\circ}$ with  respect to  the {\em minimal roughness\/}
(smallest possible edge) $R_\circ^i \propto S_i^{-1/2}$ 
of a circularly shaped domain with the same area $S_i$. 

\section{Results and Discussion}\label{Results}

Using the Eden  growth model, the KMC  procedure, and the interaction 
parameters  as described in the preceding section we present results 
for the  magnetic domain structure of a  growing ultrathin  film.  In
particular, we have determined  the average domain area and domain roughness 
as functions of the coverage  and the temperature. 
The results are always averaged over 20 runs  for systems with
same global parameters  determining  the thin film growth  and the magnetic
properties.

\begin{figure}
\begin{center}
\includegraphics*[width=5.0cm,bb=222 186 395 793,clip]
   {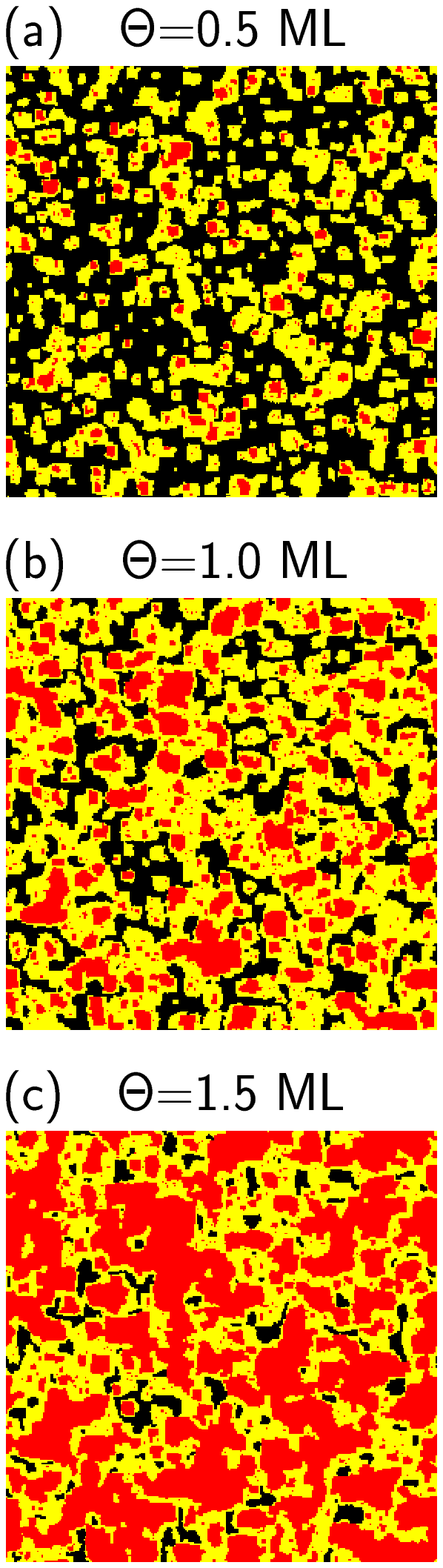} 
\end{center}
\caption{Simulation  of the  atomic structure  of a  thin film  with  a bilayer
growth mode. The unit cell has
$300\times300$  lattice constants  and contains  450  islands (island
density  $0.005$ islands  per  site).  Snapshots of three  different coverages  are
depicted,  (a)   $\Theta=0.5$~ML,   (b)  $\Theta=1.0$~ML,   and  (c)
$\Theta=1.5$~ML.  Black refers to the uncovered substrate, light 
gray to the first and dark gray to the second magnetic layer.}
\label{fig1}
\end{figure}

Examples  of  the  resulting atomic  structure  as obtained from our growth 
procedure are shown in Fig. \ref{fig1}, using the above described bilayer growth mode. 
For a better visualization of the film morphology
a unit  cell with $300\times300$ lattice sites is chosen here. 
Snapshots of three  different coverages  are
depicted,  $\Theta=0.5$~ML, $1.0$~ML,  and  $1.5$~ML. The  coverage
$\Theta=2.0$~ML  corresponds to  a smooth magnetic film with two closed 
layers (not shown).  
The resulting  atomic   structures  are  similar to those observed
experimentally for the Co/Cu(001) system \cite{ScK92}.

\begin{figure}
\begin{center}
\includegraphics*[width=13cm,bb=119 390 498 793,clip]
   {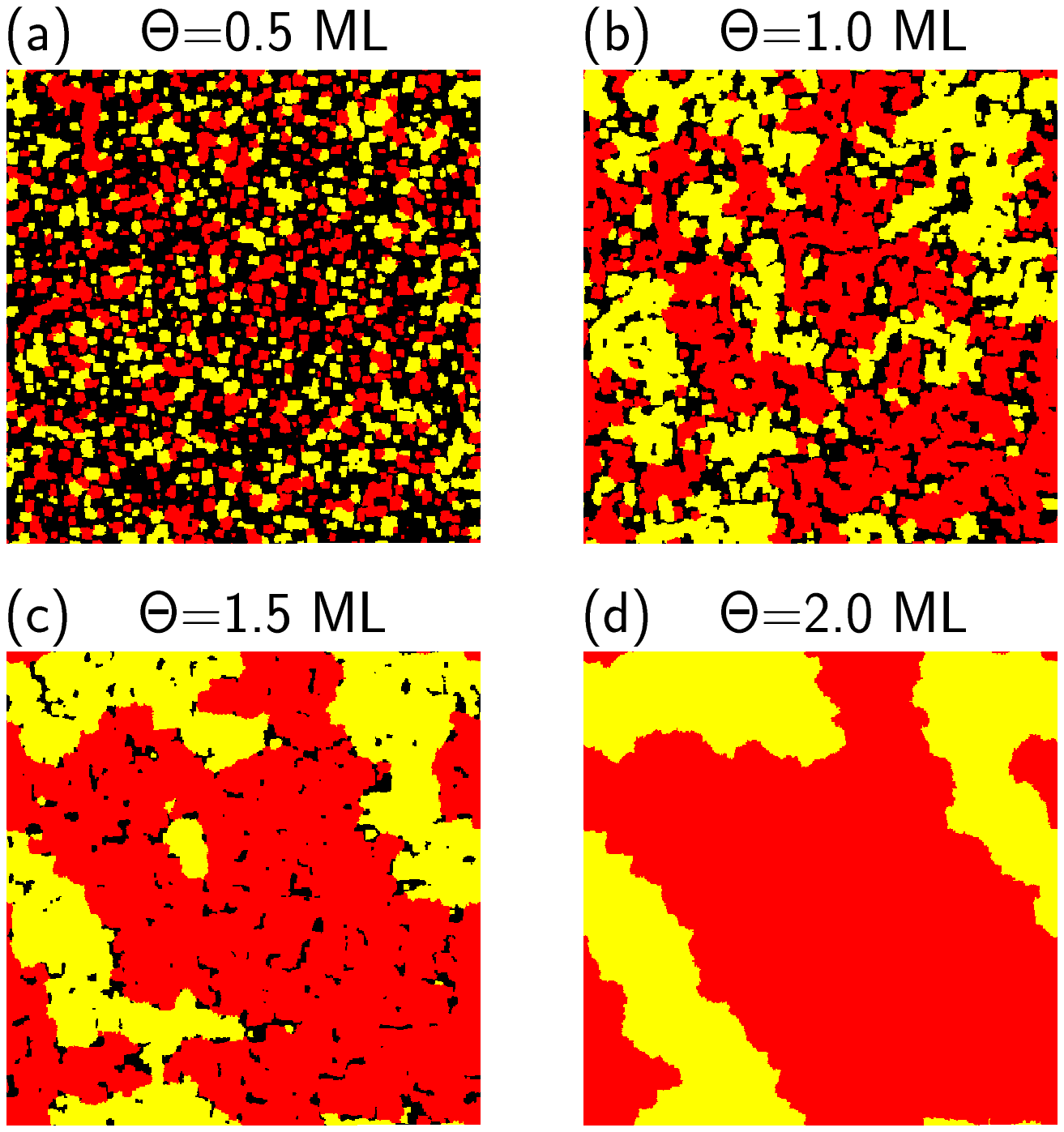} 
\end{center}
\caption{Snapshots of magnetic domain  structures  at $T=100$~K 
for  different coverages 
(a) $\Theta=0.5$~ML, (b) $\Theta=1.0$~ML, (c)  $\Theta=1.5$~ML,  and (d)
$\Theta=2.0$~ML. The unit cell has
$500\times500$  lattice constants  and contains  1250  islands (island
density  $0.005$ islands  per  site).   
Only  the  domain  pattern  of the  first  magnetic  layer  is
shown. The two gray scales refer to the two magnetic directions, 
the uncovered substrate is black.}
\label{fig2}
\end{figure}

In  Fig. \ref{fig2} we  show snapshots of resulting magnetic  domain structures  for
different coverages $\Theta$ and for the temperature $T=100$~K. 
As can  be seen, 
for small  coverages the  domain area resembles  the island  area.  For  
large  coverages $\Theta\lesssim 2$~ML the  thin film  is  
almost closed  and consequently the  magnetic domains are  very large.
Very  interestingly,  for coverages  near  and  above the  percolation
threshold  $\Theta_\mathrm{P}\sim0.9$~ML  the domains assume  an
intermediate size, covering several neighboring islands. 
We claim  that this {\em micro-domain structure\/}
is controlled and stabilized by the nonuniform atomic nanostructure of
the growing ultrathin film.  

\begin{figure}
\begin{center}
\includegraphics*[bb=108 195 507 780,clip,width=12cm]
             {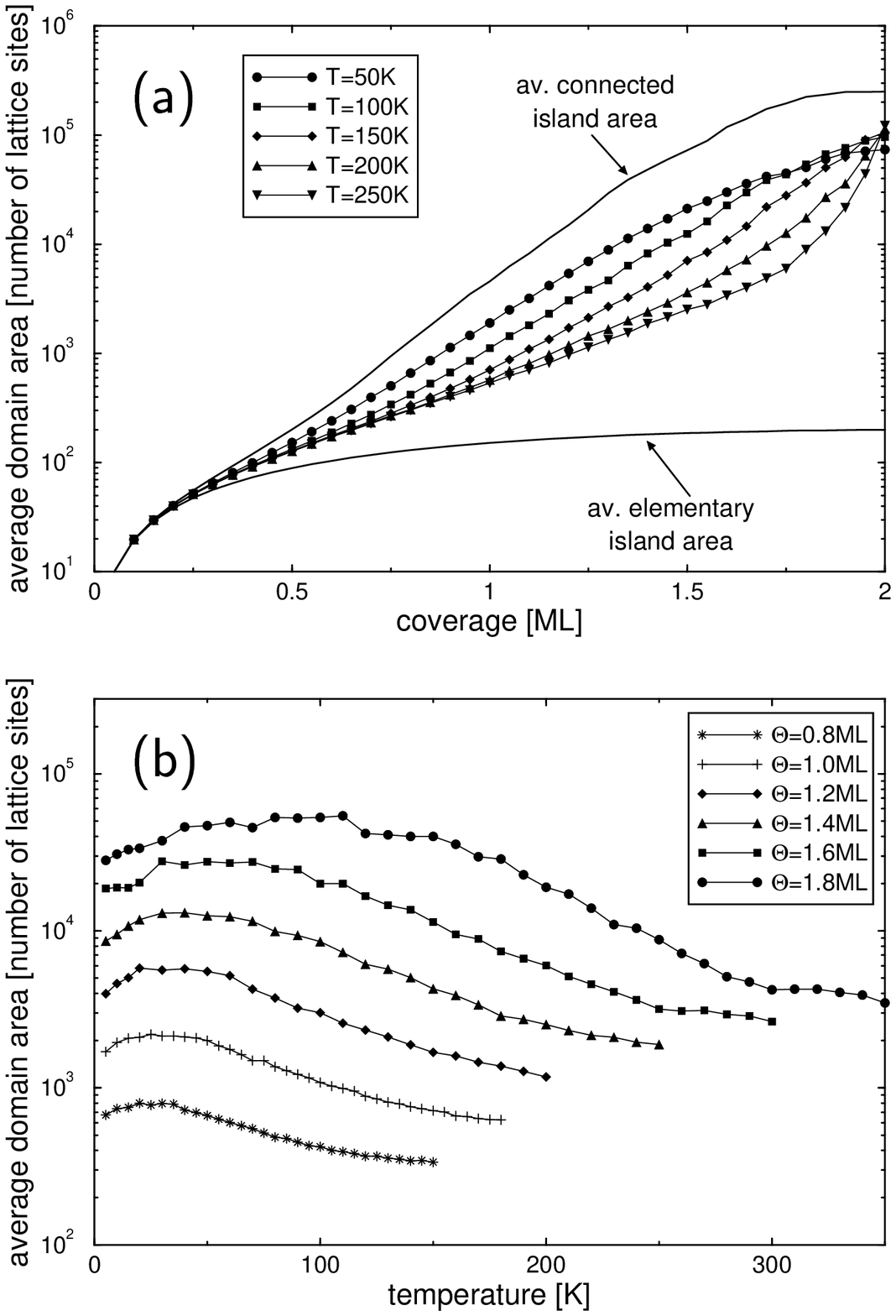}
\end{center}
\caption{Semi-logarithmic plots of the 
average domain area $\overline S$ as functions of (a) the coverage $\Theta$ 
and (b) the temperature $T$ during growth.
In (a) the average elementary 
island area and the average area of connected islands are the lower 
and upper limits of $\overline S$. 
A nanostructured thin film system with the 
atomic and magnetic properties as described in the text has been assumed. 
The results are  obtained from 20 different runs  using the same
growth mode and magnetic parameters. The area of the unit cell is 
$2.5\cdot10^5$ sites, the percolation threshold is close to 
$\Theta_\mathrm{P}\sim0.9$ ML.}
\label{fig3}
\end{figure}

From the atomic and magnetic structures 
as depicted in Figs. \ref{fig1} and \ref{fig2} we 
analyse now the average  domain  area  and  domain roughness  as functions 
of  coverage  and temperature. 
The average domain area $\overline S$ in units of lattice
sites is given in Fig. \ref{fig3}(a,b) 
as functions of the coverage  $\Theta$  and the temperature $T$. 
The average elementary island area $\overline{S}_{\mathrm{island}}$
and the average area of connected islands, which serve as lower 
and upper limits of $\overline S$, are also shown in Fig. \ref{fig3}(a).
For  coverages well  below  the percolation  threshold $\Theta_\mathrm{P}$  the
average domain area is of the order of 
$\overline{S}_{\mathrm{island}}$, see Fig. \ref{fig2}.   In this case
the ultrathin  film consists of magnetically  almost isolated islands,
the  exchange  coupling  between  neighboring  islands  has  no  large
influence.  If no additional  long range magnetic interactions such as
the  dipole coupling or the indirect exchange (RKKY-) coupling 
are  important, the  system of  isolated magnetic
islands  refers to  a superparamagnet.   Due to  their small  size and
their  superparamagnetic  behaviour   the  magnetic  domains  in  this
coverage range might be hardly visible with experimental means.

With  increasing coverage  the islands  start  to merge and form
large connected islands. Caused by the
nonuniform  island ensemble  the island coalescence  does not  occur
simultaneously in the system. In  this coverage region  the average
domain area increases considerably,  but is still markedly affected by
the atomic nanostructure  of the thin film. We  find an average domain
area definitely  larger than the  average elementary island area, but  still much
smaller than  the large domain areas for an almost closed  ferromagnetic 
film. Due to the  irregular atomic structure of the  island ensemble, 
the area of  the interfaces between neighboring  islands differ
considerably, resulting in strongly different surface contributions of
the energy barriers. Such a distribution of the effective interactions
within the island ensemble can be attributed for by a 'random exchange
energy' between neighboring island magnetizations. 
For  such a   random magnet 
metastable spin  structures exhibit considerable temporal stabilities,
characterized by large magnetic relaxation times \cite{ChB94,OhC86}.
Only after a long time the system may overcome the energy barriers and
reach its  equilibrium magnetic state.  We expect  that this mechanism
also   explains  the   appearance  and   stability  of   the  magnetic
micro-domain  structure above  the percolation  threshold of  the thin
film  as obtained  from  our calculations,  since  here the {\em irregular
atomic morphology\/} causes long  magnetic relaxation times.  

In case of a {\em  uniform\/} magnetic  island  structure, including also the 
case of a smooth ferromagnetic film, the  corresponding
relaxation times are much  shorter and a micro-domain structure might
not be observable. We have tested this assumption by simulating the 
domain structure also for a periodic array of identical magnetic islands 
with equal interactions between them, in contrast 
to an irregular island system. As 
expected, close to $\Theta_\mathrm{P}$ the domains increase fast to a large 
average size. Thus, we conclude that the frozen-in micro-domain  structure with an 
intermediate average domain area is caused by the nonuniform island ensemble. 
The experimentally observed small domains in ultrathin films with a rough 
morphology \cite{ASB90,JSO95} are expected to  
be such a metastable, nonequilibrium magnetic structure.

With further increase  of the coverage a smooth magnetic thin film is
obtained, and the domain pattern is mainly determined by competing
magnetic   interactions  (exchange,  anisotropies).  
The domain area distribution is quite wide, few large domains coexist with 
a number of small 
domains. Due to the disappearance of these small domains, see Fig. \ref{fig2}(c,d),  
the {\em average} domain  area  becomes  very  large. 
Since the extension of these large domains reaches the size of 
our unit cell, we emphasize that in the coverage range of about 1.8 -- 2~ML our 
simulation of the domain structure is influenced by finite size effects. 
As discussed in the previous section, our model should not be applied for a 
smooth film, which is present in this coverage range. 

For the temperatures 
as considered in Fig. \ref{fig3}(a), for most coverages $\Theta$ 
a larger average domain area $\overline{S}(\Theta)$ is obtained for a smaller $T$. 
Investigating $\overline{S}(T)$ as a  function of the temperature during the 
domain formation, see Fig. \ref{fig3}(b), we obtain a maximum of  $\overline{S}(T)$
in particular at low $T$. 
The reason is that for low temperatures the domain formation  is
hindered by  energy barriers.   Since with an increasing  $T$ the 
probability to overcome energy  barriers becomes larger, an increasing
average domain  area is  obtained. On the  other hand, due  to thermal
agitation a large domain  may disintegrate  into domains  with smaller
sizes, resulting  in a decreasing  average domain area  with increasing
temperatures. 
While approaching the Curie temperature $T\;\to\;T_\mathrm{C}(2{\rm ML})=350$~K 
the  average domain  area drops considerably.
With an increasing coverage  $\Theta$ the maximum of $\overline{S}(T)$ 
is shifted to larger temperatures, since the increasing average magnetic 
energy  between the islands is better suited to withstand thermal 
agitations.

\begin{figure}
\begin{center}
\includegraphics*[bb=108 206 507 783,clip,width=12cm]
             {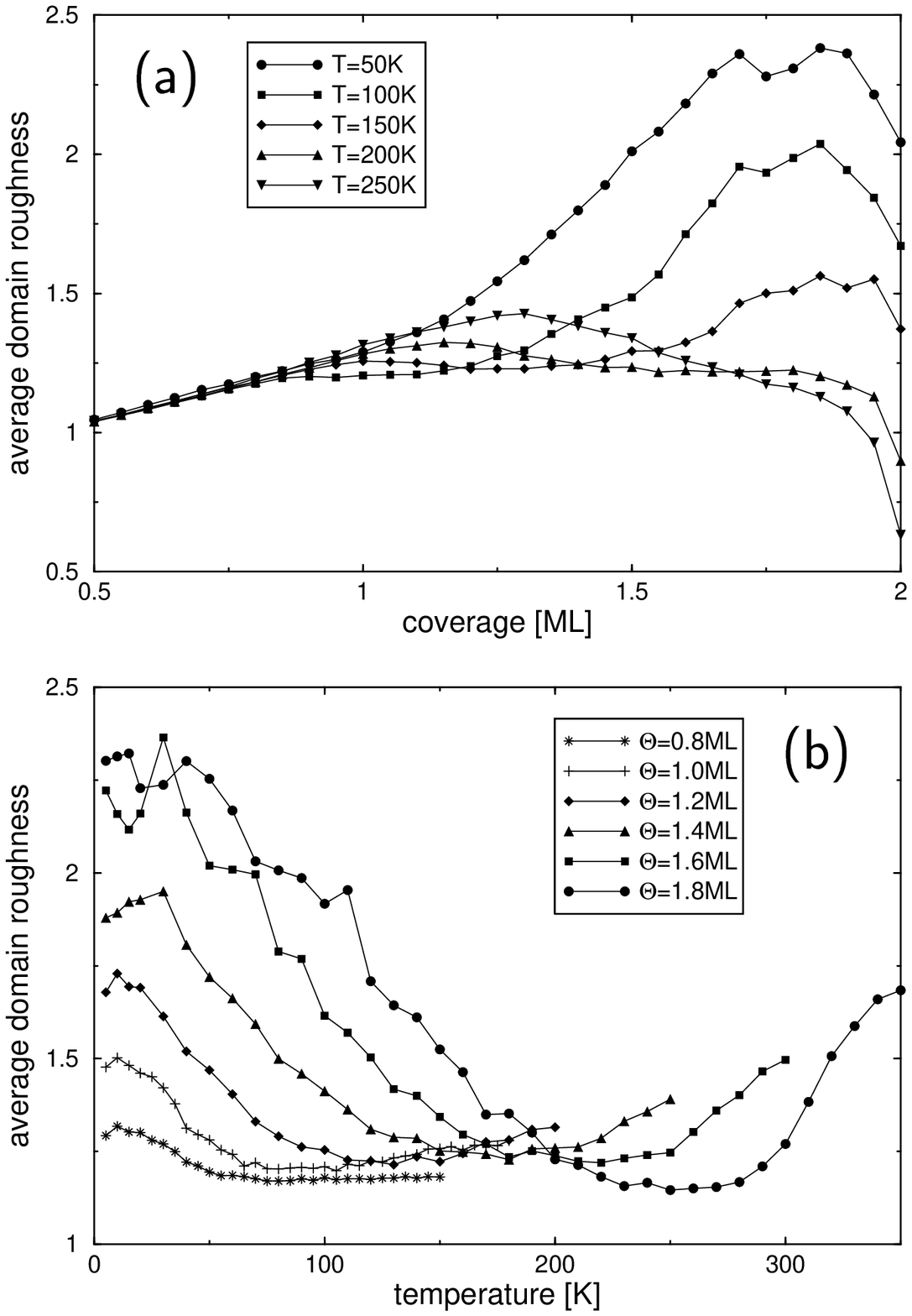}
\end{center}
\caption{Average relative domain roughness $\overline r = \overline{R/R_\circ}$
as functions of (a) the coverage $\Theta$ and (b) the temperature $T$ during growth. 
The domain roughness $R$ is given relative to the minimal
roughness $R_\circ$ of a circularly shaped  domain with  the  same 
area  as described  in Sec. \ref{Model}. 
The results are obtained from 
20 different runs using the same growth mode and magnetic parameters.
Values $\overline{r}<1$ result from single domain samples. }
\label{fig4}
\end{figure}

In Fig. \ref{fig4}(a,b) we present results for  the average domain roughness 
as  functions of the coverage $\Theta$ and the temperature $T$. 
The domain roughness is characterized by the  edge-to-area 
ratio of the domains.  As described in Sec. \ref{Model}, we present
here the {\em relative\/} 
domain roughness $\overline r=\overline{R/R_\circ}$ with respect to the 
minimal roughness $R_\circ$ of circularly shaped domains. 
Values $\overline r<1$ result from single-domain states of the simulated films. 
Similar as for the average domain area $\overline{S}(T)$, an increasing 
temperature facilitates the surmount of energy barriers, and the 
domains may assume a more compact average shape, i.e.\ are less rough. 
In contrast, an even higher temperature will destroy ordered structures 
in particular at their surface. Consequently, a minimum of 
$\overline r$ as function of $T$ is observed, see Fig. \ref{fig4}(b). 
Evidently, the relative roughness depends also on the average domain area 
and the connectivity of the system, as can be seen from Fig. \ref{fig4}(a). An 
increasing temperature tends to smoothen the domains with a large area at 
larger coverages $\Theta$.  
For lower coverages, however, the opposite behavior is observed: a 
larger roughness for an increasing temperature. This might be explained by 
the fact that in this coverage range the average domain area and the 
connectivity between the islands are comparably small so that the 
disordering effect resulting from thermal fluctuations dominates the smoothening
due to a facilitated surmount of energy barriers. 

\section{Conclusion} \label{Conclusion}
In this study we have investigated the magnetic domain structure of an
ultrathin film in the early states of film growth. 
Here the influence of the film morphology on the 
domain formation is especially strong. A model has been applied which 
is suited in particular for nanostructured systems, as well as for the 
consideration of nonequilibrium states. To our knowledge, 
the occurrence and relative stability  of the domain  structure has not 
been considered previously in connection with a growing magnetic thin 
film \cite{footnote5}. 

The average domain
area and the average  relative domain roughness have been calculated
as functions of the coverage and the temperature. In particular, we find an
interesting metastable micro-domain structure near the percolation 
threshold of the thin film.  The average area of these domains ranges 
between the
area of the elementary islands  and the  large domains  of  a smooth 
ferromagnetic  film.  We conclude that this  nonequilibrium 
micro-domain structure  is controlled and stabilized by the nonuniform  
spatial (nano-) structure of the thin film. 
Such a mechanism could lead to the unusually small domains observed in rough 
ferromagnetic films \cite{ASB90,JSO95}.  
Similar as in a random magnet a distribution of the
magnetic couplings, as present in an  irregular system, causes
large magnetic relaxation times.   For a
thin  film  system  with  a  periodic array  of  magnetic  islands
(positions,   sizes,  shapes,   interactions), a  micro-domain
structure  might not be  observable, since  large domains  will evolve
fast due to a short relaxation time.  
The average domain area is found to increase with increasing coverage,
 and to exhibit a maximum  as function of the temperature.  Similarly,
 according to  our calculations the average roughness  shows a  maximum 
as function of the coverage  and a minimum as function of 
the temperature. 

We  have considered  single-domain magnetic  islands with  a collinear
magnetization  during reversal (Stoner-Wohlfarth particles \cite{StW48}).  
This approximation is valid  if the island
size is well below the  domain wall width $w\propto\sqrt{J/K}$ 
of a bulk ferromagnet \cite{Hub98}, 
yielding an upper limit for the island size within our calculations. 
For larger sizes a noncollinear island magnetization has  to be taken 
into account \cite{HiN98}. Furthermore, 
as discussed in Sec. \ref{Model} the present model 
with nonuniform island interactions should not be applied for a 
smooth ferromagnetic film. 
Nevertheless, we expect that our approach is also suited for the investigation 
of thicker films with a considerable surface roughness. 

Also other thin film magnetic  quantities will be affected strongly by
an  irregular  nanostructure  and  the  corresponding  large  magnetic
relaxation  times. For instance,  the remanent magnetization as 
measured e.g. by MOKE,  and the magnetization reversal 
by an applied magnetic  
field may  depend considerably on the  nonuniform atomic structure \cite{SKO97}.
Since  in near  future magnetic nanostructured systems with a defined 
lateral  structure will  be prepared  in  a more  and more  controlled
manner \cite{nano}, the  corresponding magnetic  properties can be  calculated
for these  atomic structures.   Vice versa,  information about  the atomic
morphology can be obtained by analysing the magnetic properties.

Some remarks concerning the improvement of our present model study are 
in order. First, 
we  will not only  consider the two directions $S_i\pm1$ of
the magnetic islands, but will  allow for a continuous rotation of the
magnetization.   This  will result  in  particular  in a  noncollinear
magnetic arrangement  of the island  ensemble. 
The formation  of larger magnetic domains may be facilitated,  
resulting in a faster magnetic  relaxation. 
Secondly, the magnetic dipole coupling will be included in our model, 
leading possibly to an even more complex magnetic structure.
Thirdly, we will calculate explicitely the relaxational behavior of 
the magnetization for different coverages and temperatures of a 
nanostructured thin film. From an analysis of the temporal behavior 
the relaxation times can be extracted. 
Generally, also the  growth velocity  will
affect  the domain structure:  for small  velocities the  islands have
more  time to  align  themselves and  to  create larger  domains or  a
single-domain state. On the other  hand, for a fast growth the islands
become  too fast  too big  to overcome  the energy  barriers,  and the
resulting domains are expected to be smaller \cite{BJB99}.  An applied
magnetic field will support the formation of a single-domain state.
Also, we will calculate the susceptibility in response                  
to a magnetic field, which will yield additional informations for the    
interplay between the atomic and the magnetic structure of growing films.
Furthermore, the 
temperature dependence of the atomic growth can be included by applying 
appropriate growth mode parameters. In the present study we have used 
the same growth mode for all temperatures in order to concentrate 
solely on the temperature dependence of the magnetic properties. 
Finally, we  will improve our KMC method by allowing for simultaneous 
flips of connected islands (cluster-spin-flip algorithm \cite{Bin79,WaS90}). 
The consideration of such correlated jumps will yield a more 
realistic relaxational behavior.  

Acknowledgement:  This  work  has   been  supported  by  the  Deutsche
Forschungsgemeinschaft, Sonderforschungsbe\-reich 290, and by the 
Deutscher Akademi\-scher Austauschdienst (PROCOPE).

\newpage

\newpage

\newpage

\newpage

\begin{thebibliography}{00}
\bibitem{footnote2} In principle, the temperature 
dependence of the binding parameters $A(z)$ 
is approximately given by 
$A(z)\,\sqrt{q_i}\propto E^{\mathrm{B}} (q_i,z)/k_\mathrm{B} T$, 
with $E^{\mathrm{B}}(q_i,z)$ the binding energy. 
The effect of the temperature on the thin film morphology
can be indirectly taken  into  account by 
different parameters $A(z)$ and island  densities.
\bibitem{footnote3}  We emphasize that we do not consider the island 
moments to be represented by Ising spins. Within our model we 
assume that each island moment rotates continuously from one state 
 $S_i=\pm1$ to the other by passing an anisotropy dependend energy barrier. 
Thus, in the present study the anisotropy affects the magnetical dynamic.
\end{thebibliography}
\end{document}